\documentstyle[version2,aps]{revtex}
\begin{document}
\draft
\twocolumn{
\widetext
\begin{title}
Ground state properties and dynamics of the bilayer $t-J$ model
\end{title}
\author{R. Eder }
\begin{instit}
Department of Applied and Solid State Physics, University of Groningen,\\
9747AG Groningen, The Netherlands
\end{instit}
\author{Y. Ohta and S. Maekawa}
\begin{instit}
Department of Applied Physics, Nagoya University, Nagoya 464-01, Japan
\end{instit}

%                                                                      ;
\begin{abstract}
We present an exact diagonalization study of
bilayer clusters of $t-J$ model. Our results indicate a
crossover between two markedly different regimes which occurs when the
ratio $J_\perp/J$ between inter-layer and intra-layer exchange
constants increases: for small $J_\perp/J$ the data suggest the
development of $3D$
antiferromagnetic correlations without appreciable degradation of the
intra-layer spin order and the $d_{x^2-y^2}$ hole pairs within the planes
persist. For larger values of $J_\perp/J$ local singlets along the
inter-layer bonds dominate, leading to an almost complete suppression of the
intra-layer spin correlation and the breaking of the intra-layer hole pairs.
The ground state with two holes in this regime has $s$-like symmetry. The
data suggest that the crossover may occur for values of
$J_\perp/J$ as small as $0.2$. We present data for
static spin correlation function, magnetic structure factor and
spin gap.
We calculate the momentum distribution and electronic spectral
function of the "local singlet state" realized for large $J_\perp/J$, and
show that it deviates markedly from that of a single layer, making it an
implausible candidate for modelling high-temperature superconductors.
\end{abstract}

\pacs{74.20.-Z, 75.10.Jm, 75.50.Ee}

\narrowtext
\topskip12cm

%%%%%%%%%%%%%%%%%%%%%%%%%%%%%%%%%%%%%%%%%%%%%%%%%%
\section{Introduction}
%%%%%%%%%%%%%%%%%%%%%%%%%%%%%%%%%%%%%%%%%%%%%%%%%%

In connection with the so-called spin-gap phenomenon\cite{spingap}
observed in high-temperature superconductors
the properties of coupled layers of $2D$ $t-J$ model
have recently attracted considerable
attention\cite{Dagotto,MillisMonien,Lee,Normand}.
In this work we present results obtained by exact diagonalization of
small clusters, which may provide a rough guideline for this problem.
Our data were obtained by diagonalization of clusters of size
$2\times 8$ and $2\times 10$, essentially the limit that can be reached
by the exact diagonalization technique. Clearly these are still very small
systems, and subtle low-energy effects certainly will escape
our study. However, one may expect that even small clusters can mimick the
short range correlation functions (which solely determine the
total energy) roughly correct, and when interpreted with care
our calculations thus may provide a crude phase diagram of
the bilayer. Our key result is that already for fairly moderate
values of the inter-layer exchange a qualitatively new type of
ground state is realized, which deviates significantly from that
of a single layer. \\
The model under consideration reads
\[
  H =
 -\sum_{< i,j >, \sigma} t_{i,j}
( \hat{c}_{i, \sigma}^\dagger \hat{c}_{j, \sigma}  +  H.c. )
 + \sum_{< i,j >}J_{i,j}[\;\bbox{S}_i \cdot
 \bbox{S}_j
 - \frac{n_i n_j}{4}\;].
\]
Here the $\bbox{S}_i$ are the electronic spin operators,
$\hat{c}^\dagger_{i,\sigma} =c^\dagger_{i,\sigma}(1-n_{i,-\sigma})$
and the sum over $<i,j>$ stands for a summation
over all pairs of nearest neighbors. Our system consists of two
planes, labelled $A$ and $B$ for later reference,
the $z$-axis of the coordinate system is taken to be
perpendicular to the planes. We distinguish
between the hopping integral and exchange constant between nearest
neighbors in the same plane, which we denote by $t$ and $J$, and
between nearest neighbors in different planes, which we denote by
$t_\perp$ and $J_\perp$. There is presently an unclear situation
as to the correct choice of parameter values, experimental estimates
for the ratio $J_\perp/J$ vary from $0.085$\cite{RossatMignot},
over $0.3$\cite{Stern}, up to $0.55$\cite{Grueninger}.
As for the inter-layer hopping integral $t_\perp$,
the situation is even more unclear. Let us note that
inter-layer hopping integrals derived from band structure calculations
for cuprate superconductors
are meaningless on the level of the $t-J$ model, because the
"hole" in the $t-J$ model stands for an extended object, the
Zhang-Rice singlet, which has very different hybridization matrix elements
with orbitals outside the plane than a "bare" electron\cite{remark}.
We will therefore consider different ratios of the various
parameters and study their impact on e.g. the nature of the ground state.
%%%%%%%%%%%%%%%%%%%%%%%%%%
\section{Undoped bilayer}
%%%%%%%%%%%%%%%%%%%%%%%%%%
To begin with, we briefly discuss the spin correlations
of an undoped bilayer. Figure \ref{fig1}
shows the nearest-neighbor static spin correlation function
in plane and
\newpage
\topskip0cm
\noindent
out-of plane as well as the spin structure factor for
momentum tranfer
$(\pi,\pi,0)$ and $(\pi,\pi,\pi)$. As $J_\perp/J$ increases,
there is initially a fairly symmetric splitting between
$S(\pi,\pi,0)$ and $S(\pi,\pi,\pi)$, their average
$\bar{S}(\pi,\pi)=\frac{1}{2}(S(\pi,\pi,0)+S(\pi,\pi,\pi))$
remaining almost constant. The latter quantity measures the
antiferromagnetic correlations within a single plane.
In this range of $J_\perp/J$, there is virtually no
degradation of the antiferromagnetic order within the planes.
The maximum of $S(\pi,\pi,\pi)$
is reached rather precisely for $J_\perp/J=1$, and for this value
the nearest neighbor spin correlation function
within and between the planes are equal.
Quite obviously, for this parameter value the bilayer
is closest to a $3D$ Heisenberg antiferromagnet.
When $J_\perp/J$ is increased further,
$\bar{S}(\pi,\pi)$ decreases and so does the
intra-layer nearest neighbor spin correlation function.
The spin correlations
within the plane now decrease and local singlets along the
inter-layer bonds start to dominate. However, the
intra-plane correlations still are relatively "stiff", so that
large values of $J_\perp/J$ are required to bring them close to zero.\\
%%%%%%%%%%%%%%%%%%%
\section{Doped case}
%%%%%%%%%%%%%%%%%%%
We next consider the doped bilayer, more precisely
the ground states with two holes of the $2\times 8$ and $2\times 10$
bilayer. These show the same overall trends as the undoped
bilayer, however within a much smaller range of $J_\perp/J$.
Figure \ref{fig2} shows the nearest neighbor spin correlation functions
as functions of $J_\perp/J$ and demonstrates that the crossover
to a state with almost vanishing in-plane spin correlations
and dominant $z$-axis spin correlations is much more abrupt than for the
undoped system. Hole doping obviously
makes it easier for the inter-plane terms to dominate over
the in-plane exchange. The strong difference as compared to the
undoped case moreover shows that results for
undoped bilayer systems may have limited significance for the
doped bilayer.\\
The second major difference as compared to the undoped bilayer is,
that the crossover
from the small to large $J_\perp/J$ regime is now accompanied by a
ground state level crossing. For small values of $J_\perp/J$ the
$2$ hole ground state of the $2\times 10$ bilayer
belongs to the $B_1$ (or $d_{x^2-y^2}$) representation
of the $C_{4v}$ point group , reflecting the well-known
fact\cite{HasegawaPoilblanc}
that two holes in the $t-J$ model always form a bound state with
this symmetry. In this state, the two holes
can be found in the same plane with high probability
i.e. the in-plane bound state obviously persists to some degree.
This changes completely
when $J_\perp/J$ is increased beyond a certain critical value:
the ground state now has the $A_1$ (or $s$) symmetry and the two
holes are in different planes almost with probability $1$.
A rough "phase diagram" of the $2\times 10$ bilayer
is given in Figure \ref{fig3}, Table \ref{Table1}
shows the hole density correlation function
$g(\bbox{R}) = \sum_i \langle (1-n_i)(1- n_{i+\vec{R}}) \rangle $
for both $B_1$ and $A_1$ state near the crossover.
An analogous level crossing also can be found in
the $2\times 8$ cluster, for comparable values of $J_\perp/J$,
so that it does not seem to originate from some
spurious effect due to the special geometry of one of these two
clusters. Strong inter-layer spin correlations thus
are incompatible with $d$-wave pairing
of holes within the planes, and
change the symmetry properties of the ground state. It is
interesting to note that
the incompatibility of $d$-wave pairing within the planes
and strong interlayer spin correlations
has also been found by Liechtenstein {\em et al.}\cite{Mazin}
within a weak coupling framework, as well as by
Normand {\em et al.}\cite{Normand} within slave-boson mean field theory.
This seems to be a very general property of bilayer systems.\\
%%%%%%%%%%%%%%%%%%%%%%%%%%%%%%%%%%%%%%%%%%%%%%%%%%%%%%%%%%%%%%%%%%%%
At this point, the possibility of fairly strong finite-size effects
on the critical value of $J_\perp/J$ becomes obvious:
since the transition to the state with pronounced $z$-axis spin correlations
implies the breaking of the in-plane pair, it follows that an
overestimation of the in-plane hole binding energy
will lead to an artificial stabilization of the
"conventional" $B_1$ ground state.
To see this, we first define the binding energy as
$\Delta_B =
(E_0^{(2h)} -E_0^{(0h)} ) - 2\cdot (E_0^{(1h)} - E_0^{(0h)})
=(E_0^{(2h)} +E_0^{(0h)} ) - 2\cdot E_0^{(1h)}$
where $E_0^{(\nu h)}$ denotes the ground state energy with
$\nu$ holes. In small clusters of $t-J$ model this quantity is
negative\cite{HasegawaPoilblanc}, so that in the limit
$J_\perp, t_\perp \rightarrow 0$
the $2$ hole ground state of the bilayer will approach a product of an
undoped layer and a layer with two bound holes. The binding energy
$\Delta_B$ then represents a kind of charge transfer energy,
i.e. it gives the cost in energy upon transferring one
of the two holes to the other layer. Hole binding within the planes
thus opposes the formation of the $A_1$ state, and since it is known
that finite-size calculations for the $2D$ $t-J$ model tend to
substantially overestimate $\Delta_B$\cite{Bonisegni}, this may
have an impact on the crossover between the two different ground states.
In order to further investigate this issue, we add to
the Hamiltonian a repulsive interaction between holes in the same plane:
\begin{equation}
H_V = V (\sum_{i,j \in A} n_i n_j + \sum_{i,j \in B} n_i n_j ).
\label{vham}
\end{equation}
Such a term does not affect the intra-plane dynamics; it only reduces
the loss of energy which occurs when a hole is transferred between the
two layers. In particular, if we adjust $V=\Delta_B$ this term precisely
cancels the effects of the in-plane binding energy which may contain
large finite-size effects. Figure \ref{fig4} then shows
(for the value $J/t=0.4$) the change of the
crossover value of $(J_\perp/J)_{crit}$ when the in-plane repulsion is
increased. There is an obvious linear dependence and for
completely "balanced" intra-layer binding energy (i.e.
for vanishing hole transfer energy) we find
$(J_\perp/J)_{crit} = 0.17$. For $J/t=0.2$ the same
procedure gives $(J_\perp/J)_{crit} = 0.15$ (for completely balanced
binding energy), so that in the parameter region of possible relevance
for cuprate superconductors the crossover to the
unpaired state may occur for fairly small values of $J_\perp/J$.\\
For completeness we have to address possible ground state
degeneracies. For the single $10$-site cluster with periodic
boundary conditions there exists a special
permutation of the sites which leaves all nearest neighbor relations
invariant; this is analogous to the well-known $2^4$
hypercube symmetry of the $4\times 4$ cluster with periodic
boundary conditions\cite{HasegawaPoilblanc}.
This artificial symmetry operation therefore commutes with any Hamiltonian
involving only nearest neighbor terms and
introduces artificial degeneracies. Clearly, for the
$2\times 10$ bilayer there exists an analogous permutation of
$z$-axis bonds, leading to artificial degeneracies as well.
We have always neglected the degenerate ground states with finite
momentum, because we believe that the ground state
with vanishing momentum is the most natural representative
of the infinite system. In the single layer case, this is
confirmed by the fact that the $\vec{k}=0$ ground states of different
clusters all have very analogous properties, whereas the degenerate
finite momentum ground states disappear for
larger clusters. It is moreover straightforward to see, that
these degenerate states cannot add much new information:
for example the average nearest neighbor spin correlation functions
can be expressed as derivatives of the total energy
with respect to the exchange constants $J$ and $J_\perp$,
the probability to find the two holes in different planes
is related to the derivative with respect to the
parameter $V$ in (\ref{vham}), so that these quantities must be
rigorously the same for all symmetry-degenerate ground states.\\
We now turn to the spin correlations, shown in Figures \ref{fig5} and
\ref{fig6}. The $B_1$ state shows the same symmetric splitting of
$S(\pi,\pi,0)$ and $S(\pi,\pi,\pi)$ which was seen already in the
undoped spin-bilayer; the inter-plane spin correlations
increase only moderately. Quite obviously, in this state the
development of inter-layer spin correlations is accomplished
by the "lock in" of the directions of the local staggered magnetization
$M_S$ in the two planes, with essentially unchanged intra-plane
correlations. By contrast, the $A_1$ state
shows dominant inter-plane correlations and, as shown in Figure
\ref{fig2}, already for values of
$J_\perp/J\sim1$ it rapidly approaches the limit of a product state of
z-axis singlets.\\
In order to further clarify the mechanism by which
inter-layer spin correlations are built up,
we now study the probability distribution of the staggered magnetization
in the ground state. More precisely, we denote by
${\cal M}_{\mu,\nu}$ the set of basis states
which have staggered magnetization $\mu$ in the A-plane and staggered
magnetization $\nu$ in the B-plane.
Then, we can define the operator
\[
{\cal P}_{\mu,\nu} = \sum_{\alpha \in {\cal M}_{\mu,\nu}}
| \Phi_\alpha \rangle \langle \Phi_\alpha |,
\]
and its ground state expectation value
$S(\mu,\nu) = \langle {\cal P}_{\mu,\nu} \rangle$.
It may be interpreted as the total weight of states
with single-layer staggered magnetizations $\mu$ and $\nu$.
Figure \ref{fig7} shows this quantity for two values of
$J_\perp/J$. For vanishing inter-layer coupling the ground state
is simply the product of two single-layer ground states, and if one
chooses fixed $z$-component of the momentum,
$S(\mu,\nu)$ must be completely symmetric, i.e.
$S(\mu,\nu)=S(\mu,-\nu)=S(-\mu,\nu)=S(-\mu,-\nu)$.
The data then show, that for increasing
$J_\perp/J$ there is first  a slight shift of weight from the
$(+,+)$ and $(-,-)$ quadrants to the $(+,-)$ and $(-,+)$
quadrants: the staggered magnetization of the two layers "locks in".
However, states with large staggered magnetization
still do have appreciable weight, and the correlation between the
values of $M_S$ in the two planes is not yet strong.
This changes completely when we increase $J_\perp/J$ beyond the
crossover to the $A_1$ ground state:
$S(\mu,\nu)$ is now concentrated near the line $\mu=-\nu$,
and along this line it drops sharply towards large values
of $M_S$. The values of $M_S$
in the two layers are now strongly correlated: when there
is e.g. a fluctuation towards a large positive value of $M_S$ in the
$A$-plane, it must be accompanied by a fluctuation to a large negative
value in the $B$-plane. These results are precisely what one would
expect if the spins were distributed randomly within the planes,
but under the constraint that nearest neighbor spins in
$z$-direction always be antiparallel.
Analysis of the probability distribution
$S_{\mu,\nu}$ thus confirms the scenario
inferred from the spin correlation function: there is a regime of small
$J_\perp/J$, where the system responds to the
inter-layer coupling by a weak "lock-in" of the staggered
magnetization in the two planes, while no appreciable change of the
intra-layer correlations takes place. For large $J_\perp/J$, on the
other hand, we
have the clear signatures of the $z$-axis singlet state.\\
%%%%%%%%%%%%%%%%%%%%%%%%%%%%%%%%%%%%%%%%
We next consider the "spin gap", defined as
\[
\Delta_{spin} = E_{0,\Gamma_0}(\pi,\pi,\pi) - E_0
\]
Here $E_0$ denotes the energy of the ground state
(which has momentum $0$) and $E_{\Gamma_0}(\pi,\pi,\pi)$
denotes the lowest energy in the subspace of states with momentum
$(\pi,\pi,\pi)$ and the same point group symmetry $\Gamma_0$ as the
ground state.
Inspection shows that the $\vec{k}=0$ ground state is always a singlet,
the $(\pi,\pi,\pi)$ state always a triplet, so that $\Delta_{spin}$ indeed
represents the gap which would be observed in the dynamical spin correlation
function with momentum transfer $(\pi,\pi,\pi)$. We note that
$\Delta_{spin}$
is prone to massive finite-size effects, as can be seen e.g. from
the fact that even an undoped $2D$ cluster has a "spin gap" of
$\approx 0.4J$\cite{remark2}. The values of $\Delta_{spin}$ determined
from the small
clusters thus may at best indicate a rough qualitative trend, but
clearly have no quantitative significance.
Then, Figure \ref{fig8}a shows the spin gap over a wide range of
$J_\perp/J$.
It is remarkable that the gap initially {\em decreases} with increasing
$J_\perp$ and in fact shrinks appreciably at the level crossing to
the strong $z$-axis correlation ($A_1$) state. It is only for values of
$J_\perp/J>1$ that the spin gap becomes large and grows linearly
with $J_\perp$, reflecting the saturation of the
$z$-axis spin correlation function (see Figure \ref{fig2}).
For an undoped system and
in the limit $J_\perp/J\rightarrow \infty$ it is easy to see that
a single $z$-axis {\em triplet} bond can propagate in a "background" of
$z$-axis singlets with a nearest neighbor hopping element of
$J/2$, so that the spin gap should be $J_\perp - z \cdot (J/2)$,
with $z$ the number of nearest neighbor $z$-axis bonds.
The resulting asymptotic expressions $\Delta_{spin} = J_\perp - J$
for the ladder ($z=2$) and $\Delta_{spin} = J_\perp -2J$
for the bilayer ($z=4$) indeed provide excellent estimates for the
undoped systems (compare Figure 2 in Ref. \cite{Dagotto}); for the
doped system, however, the numerical result rather suggests
the asymptotic expression $\Delta_{spin} = J_\perp - (J/2)$, which
would imply a reduction of the effective hopping element for the
triplet-bond by a factor of $4$.
This indicates, that the propagation of the $z$-axis triplet
is inhibited by the presence of mobile holes, so that
there must be a strong interplay between the local singlets/triplets
and the holes.\\
Figure \ref{fig8}b shows the spin gap for other parameter values in the
range $J_\perp/J<1$ and demonstrate that increasing values of $J_\perp/J$
initially tend to reduce $\Delta_{spin}$, as would be expected from an
enhancement of antiferromagnetic correlations.
It is interesting to note that the crossover $B_1 \rightarrow A_1$
never leads to an increase of $\Delta_{spin}$.
While the significance of these data should not be
overestimated, we may state that in the small clusters
there is no indication of any
widening of the spin gap due to interlayer coupling, at least not for
"reasonable" values of $J_\perp/J$.\\
%%%%%%%%%%%%%%%%%%%%%%%%%%%%%%%%%%%%%%%%%%%%%%%%%
\section{Local singlet state}
%%%%%%%%%%%%%%%%%%%%%%%%%%%%%%%%%%%%%%%%%%%%%%%%%
We now proceed to a study of the "$z$-axis singlet state", which is
realized for larger $J_\perp/J$. As has been shown above,
this state may appear already for quite moderate values of
$J_\perp/J\approx 0.2$, which have in fact been postulated to
be realized in actual high-temperature superconductors. In fact,
this state may be considered a qualitatively new feature of the
bilayer system, as compared to the single-layer ground states
studied extensively before. We first consider the limiting case
of vanishing in-plane parameters $J$ and $t$\cite{Dagotto}.
Depending on the relative
magnitude of $t_\perp$ and $J_\perp$, the ground state
of the $2 \times N$ bilayer
with an even number of holes $n_h$ can take two quite different
forms (see Figure \ref{fig9}): it is either the product
of $N- n_h/2$ $z$-axis singlets and $n_h/2$ empty $z$-axis bonds
with energy $(N-n_h/2) \cdot J_\perp$; or it is a product of
$N-n_h$ $z$-axis singlets and $n_h$ singly occupied
$z$-axis bonds with energy $(N - n_h)\cdot J_\perp - n_h \cdot t_\perp$.
The degeneracy of this ground state is
lifted by the intra-plane terms of the Hamiltonian.
We assume that the most
important intra-plane term in the Hamiltonian is the
hopping integral $t$. Then, a singly occupied bond can propagate
to a nearest neighbor
in a single-step process with the effective hopping integral $t/2$,
whereas an empty bond propagates to a nearest neighbor
via a two step process
where the intermediate state has a broken $z$-axis singlet;
in the regime of large $J_\perp/t$ perturbation theory
would give an effective hopping integral $\sim t^2/J_\perp$.
We thus have the interesting situation that, depending on the
relative strength of parameters, two very different "fixed points" may
be appropriate to describe the bilayer system:
if the arrangement of holes in empty bonds is preferred, the system
should correspond to a gas of $n_h/2$ hard core bosons, propagating
via nearest neighbor hopping of strength $\sim t^2/J_\perp$.
On the other hand, if the arrangement of holes along
singly occupied bonds is preferred, the bilayer should correspond to
a system of $n_h$ spin-$1/2$ fermions, propagating via nearest neighbor
hopping of strength $t/2$. In both cases, however, there is a positive
nearest neighbor hopping matrix element for either  hole-like "Fermion" or
"Boson", so that the holes may be expected to accumulate
at the in-plane momentum $(\pi,\pi)$.\\
The calculated values of the electron momentum distribution function
$n(\vec{k})= \langle \hat{c}_{\vec{k},\sigma}^\dagger
\hat{c}_{\vec{k},\sigma}\rangle$, given in Tables \ref{Table2} and
\ref{Table3}, then indeed show an
enhanced hole-occupancy of $(\pi,\pi)$ in the $A_1$ state even
for moderate values of $J_\perp$ and $t_\perp$:
this is very pronounced in the $2\times 8$ system, but also in the
$2\times 10$ bilayer one can realize an enhanced depression
at $(\pi,\pi)$ and a tendency towards the "flattening"
of $n(\vec{k})$ in the remainder of the Brillouin zone.
This indicates that the momentum distribution with
"hole pockets" at $(\pi,\pi)$, which would be expected in the
limit of large inter-layer coupling indeed seems to persist
in the $A_1$ state also for moderate values of the
inter-layer coupling. This picture is confirmed by the
single particle spectral function $A(\bbox{k},\omega)$, shown in
Figure \ref{fig10}. Due to restrictions in computer memory, we
have evaluated this quantity only for the $2\times 8$ bilayer.
The topmost peaks in the photoemission spectra are located
at $(\pi/2,\pi/2)$ and $(\pi,0)$ (due to the special
symmetry of the $2D$ $8$-site cluster these momenta are in fact
identical for the $A_1$ state), but there is no appreciable
low energy weight in the inverse photemission spectrum at either of these
momenta: quite obviously these $k$-points are "occupied".
This is in clear contrast to the spectral function for the
$B_1$ state\cite{BilayerI} which showed a rather obvious Fermi level
crossing at $(\pi,0)$. Another notable feature is the strong
contrast between the relatively sharp peaks in the
bonding channel (i.e. $k_z=0$) and the more diffuse spectra in the
antibonding one, which have markedly enhanced incoherent continua.
This indicates that electrons in the
bonding channel are the "more well-defined" excitations
of the system, as one would expect it for a product state
of doubly and singly occupied bonds.
The strong difference between the $k_z=0$ and $k_z=\pi$ spectra
is again in contrast to
their almost identical shape in the $B_1$ ground state\cite{BilayerI}.
The available data thus rather consistently suggest that
even for moderate values of $J_\perp$ and $t_\perp$ the
"$z$-axis singlet state" bears a strong resemblance to the
$J_\perp, t_\perp \rightarrow \infty$ limit, which in particular implies
the accumulation of holes at $(\pi,\pi)$.\\
%%%%%%%%%%%%%%%%%%%%%%%%%%%%%%%
\section{Summary and Discussion}
%%%%%%%%%%%%%%%%%%%%%%%%%%%%%%
In summary, we have studied the effects of interlayer exchange in a
bilayer $t-J$ model. The data rather consistently suggest
a crossover between two very different regimes:
for small and moderate values of the inter-plane exchange coupling
there is no qualitative change of the spin correlations, the
in-plane hole pairing with $d_{x^2-y^2}$ symmetry persists
and the inter-plane exchange induces $3D$ antiferromagnetic correlations.
For increasing values of $J_\perp/J$ there is a crossover to a state
with dominant
inter-layer spin correlations, where the intra-layer hole pairing
is completely suppressed.
We note that this feature may well persist in the infinite system if the
pairing state remains short ranged; in this case, it would be the short
range corrrelations which dominate the physics, and these may be expected
to be described properly even in the small clusters.
The data finally do not show any
indication of a widening of the spin gap at the crossover.\\
The result that strong inter-layer coupling suppresses the in-plane
$d_{x^2-y^2}$ pairing clearly would be hard to reconcile with
the ever growing experimental evidence for $d$-type symmetry of the
superconducting order parameter
in the cuprate superconductors\cite{Wollmann} and
the experimental fact that  $T_c$ seems to increase with the number of
closely coupled layers in these compounds. In addition, our data indicate
that the "$z$-axis singlet state" realized for large $J_\perp/J$
has fairly unusual properties such
as a tendency towards "hole pockets" at the in-plane momentum $(\pi,\pi)$,
so that it does not seem to be a very promising candidate for modelling
real high-temperature superconductors.
Moreover, the relatively strong difference between the
$z$-axis singlet state and the "conventional" single layer ground state
would make it hard to understand why many properties of single and
double layer cuprates are relatively similar.\\
Clearly, the simplest way to resolve these problems is to assume that
the values of $J_\perp/J$ in the actual materials are too small
to ever induce the $z$-axis singlet state. Unfortunately
strong finite size effects do not allow us to
give a very accurate estimate of the critical value
where the crossover occurs.
Then, a possible enhancement of antiferromagnetic
spin correlations due to the onset of $3D$ correlations
in the more conventional ground states realized for lower $J_\perp/J$
may be more easy to reconcile with experiment: it seems
plausible that pairing theories which rely on
antiferromagnetism, such as the antiferromagnetic
spin fluctuation theory\cite{Moriya}
or the antiferromagnetic van-Hove scenario\cite{Dagottoetal}
would profit from
the onset of $3D$ correlations, so that the exchange coupling of several
$Cu O_2$ planes could be favourable for achieving higher $T_c$'s.\\

We thank Dr. W. Koshibae for help with preparing the manuscript.
Financial support of  R. E. at Nagoya by the Japan Society for the
Promotion of Science is most gratefully acknowledged. Support of
R. E. at Groningen by the European Community is
most gratefully acknowledged, too.\\[3cm]

%%%%%%%%%%%%%%%%%%%%%%%%%%%%%%%%%%%%%%%%%%%%%%%%%%%%%%%%%%%%%%%%
\begin{table}
\caption{Comparison of the hole density correlation function
$g(\vec{R})$ in the lowest state with $B_1$ symmetry
and the lowest state with $A_1$ symmetry of the
$2\times 10$ bilayer $t-J$ model.
Parameter values are $t=1$, $J=0.4$, $t_\perp=0.5$, $J_\perp/J=0.8$.
The value of $J_\perp/J$ is chosen slightly above the
crossover $B_1 \rightarrow A_1$, so that the $A_1$ state is the
ground state, the $B_1$ state being $0.0289$ above the ground state.}
\begin{tabular}{l| c c c c}
$\;$ & $\;$ & $\bbox{R}_{in-plane}$ & $\;$ & $\;$ \\
\hline
$B_1$ & $(0,0)$ & $(1,0)$ & $(1,1)$ & $(2,1)$ \\
\hline
$R_z=0$ & 0.000 & 0.0791 & 0.0943 & 0.0898 \\
$R_z=1$ & 0.0189 & 0.0243 & 0.0213 & 0.0151 \\
\hline
$A_1$ & $(0,0)$ & $(1,0)$ & $(1,1)$ & $(2,0)$ \\
\hline
$R_z=0$ & 0.000 & 0.0115 & 0.0095 & 0.0119 \\
$R_z=1$ & 0.2217 & 0.1038 & 0.0573 & 0.0380 \\
\end{tabular}
\label{Table1}
\end{table}
%%%%%%%%%%%%%%%%%%%%%%%%%%%%%%%%%%%%%%%%%%%%%%%%%%%%%%%%%%%%%%
\begin{table}
\caption{Comparison of the
momentum distribution function $n(\vec{k})$ for the
lowest state with $A_1$ symmetry and the lowest state with
$B_1$ symmetry of the $2 \times 10$ bilayer $t-J$ model.
Parameter values are like in Table 1. }
\begin{tabular}{l| c c c c}
$\;$ & $\;$ & $\bbox{k}_{in-plane}$ & $\;$ & $\;$ \\
\hline
$A_1$ & $(0,0)$ & $(\frac{3\pi}{5},\frac{\pi}{5})$ &
$(\frac{2\pi}{5},\frac{4\pi}{5})$ & $(\pi,\pi)$ \\
\hline
$k_z=0$ & 0.5357 & 0.5321 & 0.4367 & 0.2757\\
$k_z=\pi$ & 0.5214 & 0.4730 & 0.3966 & 0.3132 \\
\hline
$B_1$ & $(0,0)$ & $(\frac{3\pi}{5},\frac{\pi}{5})$ &
$(\frac{2\pi}{5},\frac{4\pi}{5})$ & $(\pi,\pi)$ \\
\hline
$k_z=0$ & 0.5399 & 0.5278 & 0.4374 & 0.3424\\
$k_z=\pi$ & 0.5416 & 0.4815 & 0.3668 & 0.3223 \\
\end{tabular}
\label{Table2}
\end{table}
%%%%%%%%%%%%%%%%%%%%%%%%%%%%%%%%%%%%%%%%%%%%%%%%%%
\begin{table}
\caption{Comparison of the
momentum distribution function $n(\vec{k})$ for the
lowest state with $A_1$ symmetry and the lowest state with
$B_1$ symmetry of the $2 \times 8$ bilayer $t-J$ model.
The value of $J_\perp/J=0.6$ is chosen slightly above the
crossover $B_1 \rightarrow A_1$, so that the $A_1$ state is the
ground state, the $B_1$ state being $0.0154$ above the ground state.
The other parameter values are $t=1$, $J=0.4$, $t_\perp=0.5$.}
\begin{tabular}{l| c c c c}
$\;$ & $\;$ & $\bbox{k}_{in-plane}$ & $\;$ & $\;$ \\
\hline
$A_1$ & $(0,0)$ & $(\frac{3\pi}{5},\frac{\pi}{5})$ &
$(\frac{2\pi}{5},\frac{4\pi}{5})$ & $(\pi,\pi)$ \\
\hline
$k_z=0$ & 0.5446 & 0.5014 & 0.5014 & 0.16574\\
$k_z=\pi$ & 0.5017 & 0.4213 & 0.4213 & 0.2518 \\
\hline
$B_1$ & $(0,0)$ & $(\frac{3\pi}{5},\frac{\pi}{5})$ &
$(\frac{2\pi}{5},\frac{4\pi}{5})$ & $(\pi,\pi)$ \\
\hline
$k_z=0$ & 0.5469 & 0.5260 & 0.4410 & 0.2915\\
$k_z=\pi$ & 0.5444 & 0.4624 & 0.2525 & 0.2761 \\
\end{tabular}
\label{Table3}
\end{table}

\figure{(a) Static nearest neighbor spin correlation function
        $\langle \vec{S}_i \cdot \vec{S}_j \rangle$
        for the ground state of the undoped $2\times 10$ bilayer.
        (b) Static spin structure factors
        $S(\pi,\pi,0)$ (diamonds) and $S(\pi,\pi,\pi)$ (crosses)
        and average $\bar{S}(\pi,\pi)$ (squares) for the same
        state.
\label{fig1}}

\figure{Static nearest neighbor spin correlation function
        $\langle \vec{S}_i \cdot \vec{S}_j \rangle$
        for the ground state of the $2\times 10$ bilayer with two holes.
        Parameter values are $t=1$, $J=0.2$, $t_\perp=0.1$.
\label{fig2}}

\figure{"Phase diagram" for the $2\times 10$ bilayer
        with two holes for two values of $t_\perp/t$.
        The figure indicates the regions of stability
        of the $A_1$ and $B_1$ ground state for different
        parameter values. The respective critical values of
        $J_\perp/J$ have been
        obtained by interpolating ground state energies
        and are accurate only to $\sim \pm 0.05$.
\label{fig3}}
\figure{Critical value of $(J_\perp/J)_{crit}$ for the crossing
        $B_1 \rightarrow A_1$
        as a function of the intra-plane repulsion parameter $V$.
\label{fig4}}
\figure{Nearest neighbor spin correlation function
        $\langle \vec{S}_i \cdot \vec{S}_j \rangle$
        for the $2\times 10$ bilayer with two holes.
        (b) Static spin structure factors
        $S(\pi,\pi,0)$ (diamonds) and $S(\pi,\pi,\pi)$ (crosses)
        and average $\bar{S}(\pi,\pi)$ (squares) for the same system.
        Parameter values are $t=1$, $J=0.4$, $t_\perp=0.5$.
        \label{fig5}}
\figure{Same as Figure \ref{fig4} for $J=0.2$.
        \label{fig6}}
\figure{Probability distribution $S_{\mu,\nu}$ for
        the ground state of the $2\times 8$ bilayer with two
        holes and different $J_\perp/J$. The distribution has been
        averaged over the areas indicated in the figure. Parameter values
        are $J_\perp/J=0.4$ (a) and
        $J_\perp/J=1$ (b), the other parameters are
        $t=1$, $J=0.4$, $t_\perp=0.5$ for both figures.
\label{fig7}}
\figure{(a) Spin gap $\Delta_{spin}$ as a function of $J_\perp/J$ for
        $J=0.2$, $t_\perp=0.1$, $2\times 10$ bilayer.\\
         (b) $\Delta_{spin}$ for different parameter values in the
         $2 \times 10$ bilayer, $t=1$.
\label{fig8}}
\figure{Possible ground states of the bilayer model in the limit
        $t,J \rightarrow 0$: the holes are on "singly occupied bonds"
        (a) or on "empty bonds" (b).
\label{fig9} }
\figure{Single particle spectral function $A(\bbox{k},\omega)$
        for the $2\times 8$ bilayer. The vertical line markes the
        Fermi energy, the frquency range $\omega>E_F$ ($\omega<E_F$)
        corresponds to electron creation (annihilation).
        The full (dashed) line refers to the bonding (antibonding) channel.
        Parameter values are like in Table \ref{Table3}.
\label{fig10} }
%%%%%%%%%%%%%%%%%%%%%%%%%%refs%%%%%%%%%%%%%%%%%%%%%%%%%%%%%%%%%%%%%%

}

\begin{references}
\bibitem{spingap}
  H. Yasuoka {\it et al.}, in
  {\em Strong Correlations and Superconductivity},
  Eds. H. Fukuyama, S. Maekawa, and A. P. Malozemoff,
  Springer Series in Solid State Sciences {\bf 89}, 254, (1989).
\bibitem{Dagotto}
  E. Dagotto, J. Riera, and D. Scalapino,
  Phys. Rev. B {\bf 45}, 5744 (1992).
\bibitem{MillisMonien}
  A. J. Millis and H. Monien, Phys.\ Rev.\ Lett.\
  {\bf 70}, 2810 (1993); Phys. Rev. B {\bf 50}, 16606 (1994).
\bibitem{Lee}
  M. Ubbens and P.A. Lee, Phys. Rev. B {\bf 50}, 438 (1994);
  K. Kuboki and P.A. Lee, preprint.
\bibitem{Normand}
  B. Normand, H. Kohno, and H. Fukuyama, preprint.
\bibitem{RossatMignot}
  J.Rossat-Mignot, {\it et al}, in {\em Dynamics of Magnetic fluctuations
  in High-Temperature Superconductors}, Eds. G. Reiter, P. Horsch and
  G. C. Psaltakis, NATO-ASI Series B, Vol. 246, 35 (1991).
\bibitem{Stern}
  R. Stern {\em at al.}, preprint.
\bibitem{Grueninger}
  M. Gr\"uninger {\em et al.}, SISSA preprint.
\bibitem{remark}
  For example it is straightforward to see that due to its nontrivial
  symmetry the Zhang-Rice singlet on a given plaquette has vanishing
  hybridization with both the $Cu$ $4s$ orbital on the
  central $Cu$ atom and with any $p$ orbital of the apex oxygen.
  $Z$-axis propagation is only possible via orbitals in neighboring
  plaquettes.
\bibitem{HasegawaPoilblanc}
  Y. Hasegawa, D. Poilblanc,
  Phys.\ Rev.\ B {\bf 40}, 9035 (1989).
\bibitem{Mazin}
  A. I. Liechtenstein, I. I. Mazin, and O. K. Andersen,
  Phys. Rev. Lett. {\bf 74}, 2303 (1995).
\bibitem{Bonisegni}
  M. Bonisegni and E. Manousakis, Phys. Rev. B {\bf 43}, 10353 (1992).
\bibitem{remark2}
  More precisely, the values are $0.329J$ for $16$ sites,   $0.506J$ for
  $18$ sites and $0.467J$ for $20$ sites.
\bibitem{BilayerI}
  R. Eder, Y. Ohta, and S. Maekawa, Phys. Rev. B {\bf 51}, 3265 (1995).
\bibitem{Wollmann}
  D.A. Wollman {\it et al.}, Phys. Rev. Lett. {\bf 71}, 2134 (1993);
  J. R. Kirtley {\it et al.}, Nature {\bf 373}, 225 (1995).
\bibitem{Moriya}
  T. Moriya, Y. Takahashi, and K. Ueda,
  J. Phys. Soc. Jpn. {\bf 49}, 2905 (1990);
  P. Monthoux, A. Balatsky, and D. Pines,
  Phys.\ Rev.\ Lett. {\bf 67}, 3448 (1991).
\bibitem{Dagottoetal}
E. Dagotto {\em et al.}
Phys.\ Rev.\ Lett. {\bf 74}, 310 (1995).

\end{references}
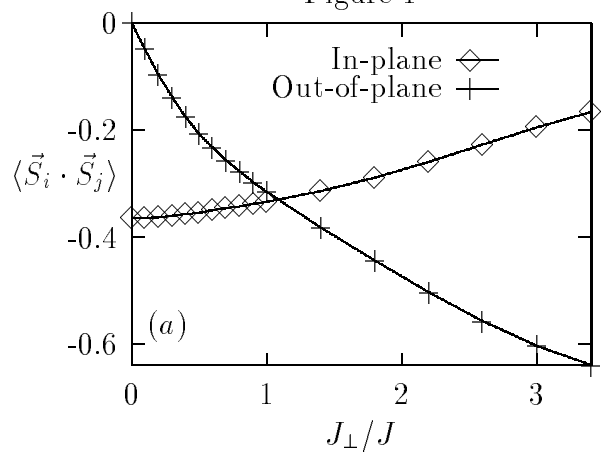
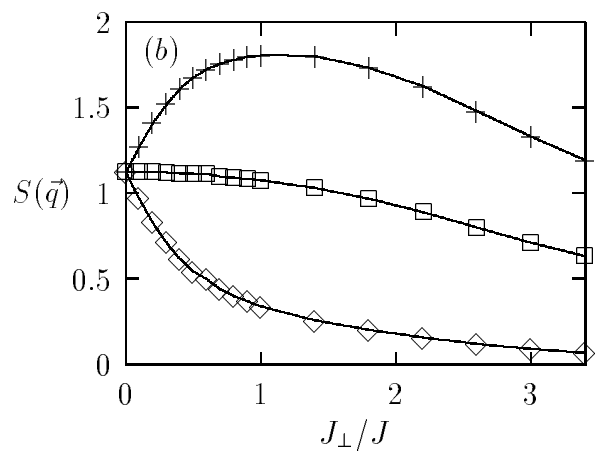
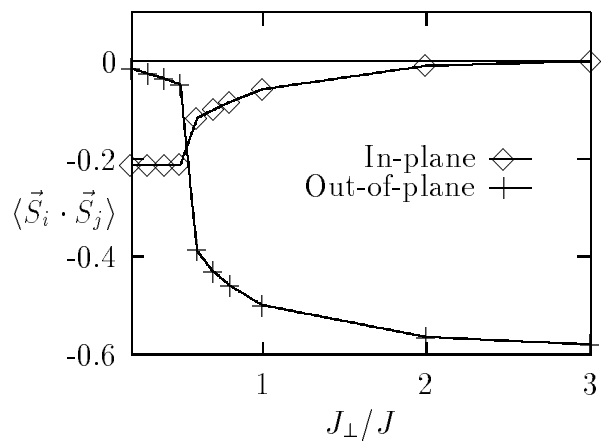
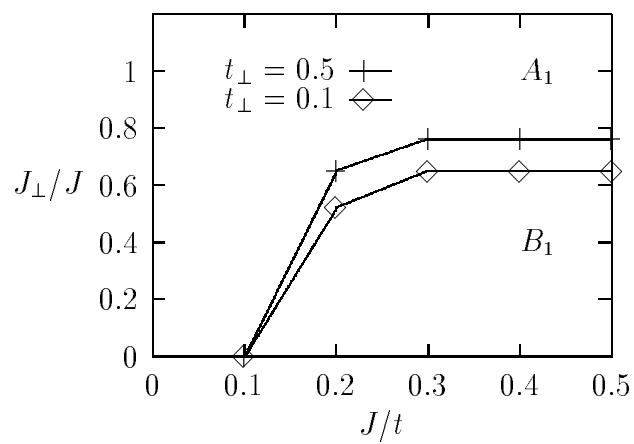
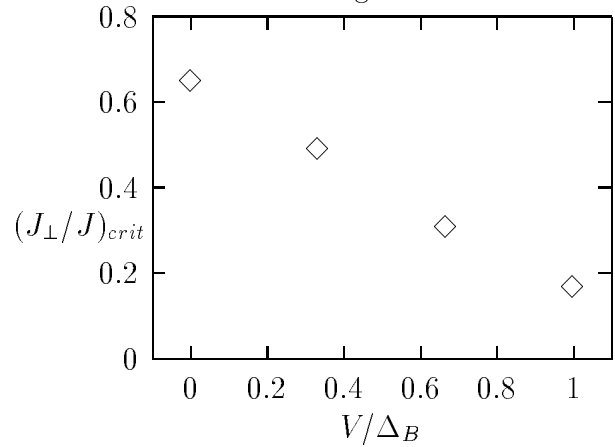
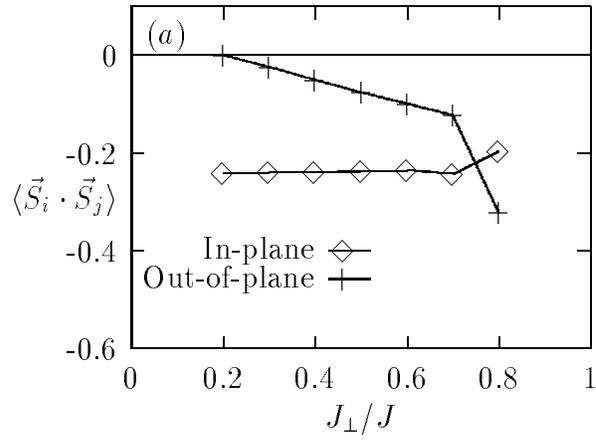
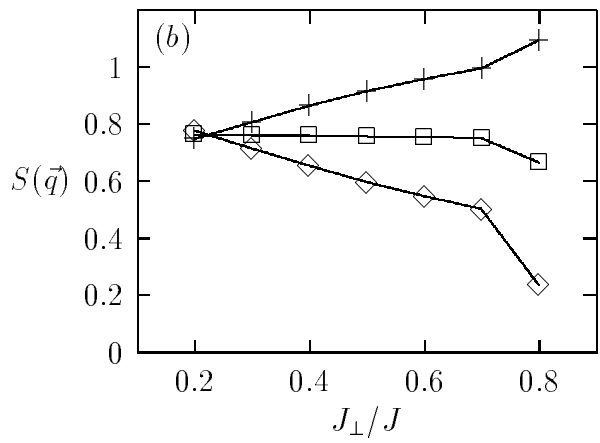
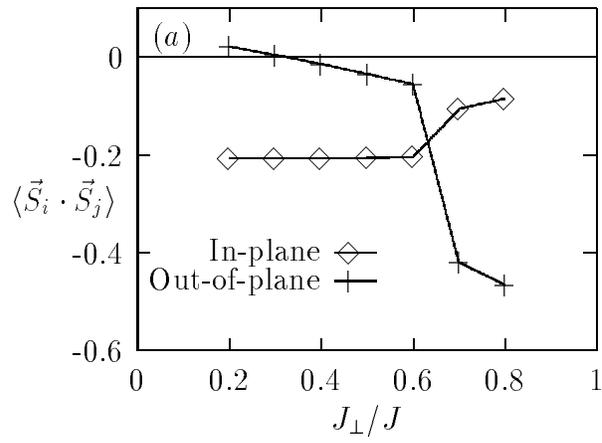
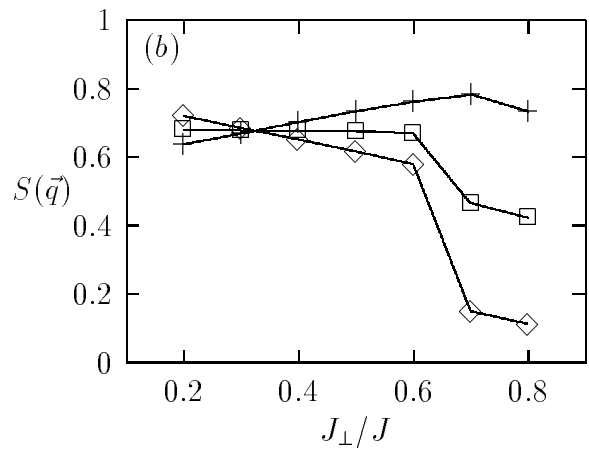
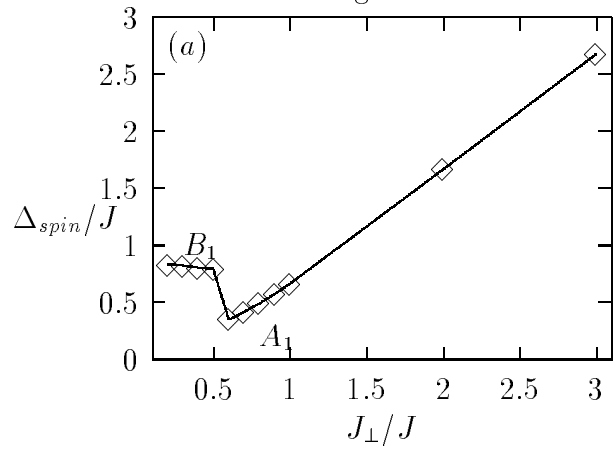
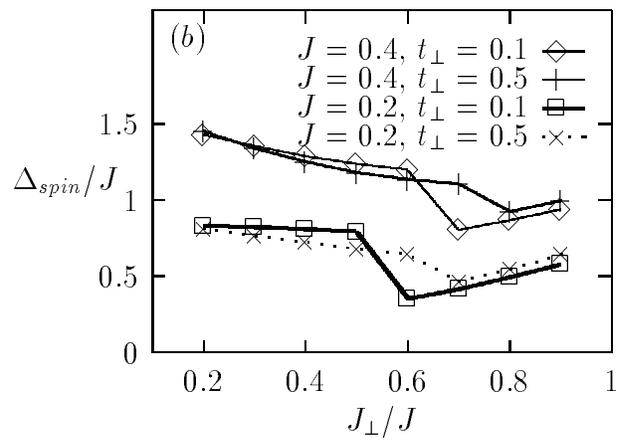
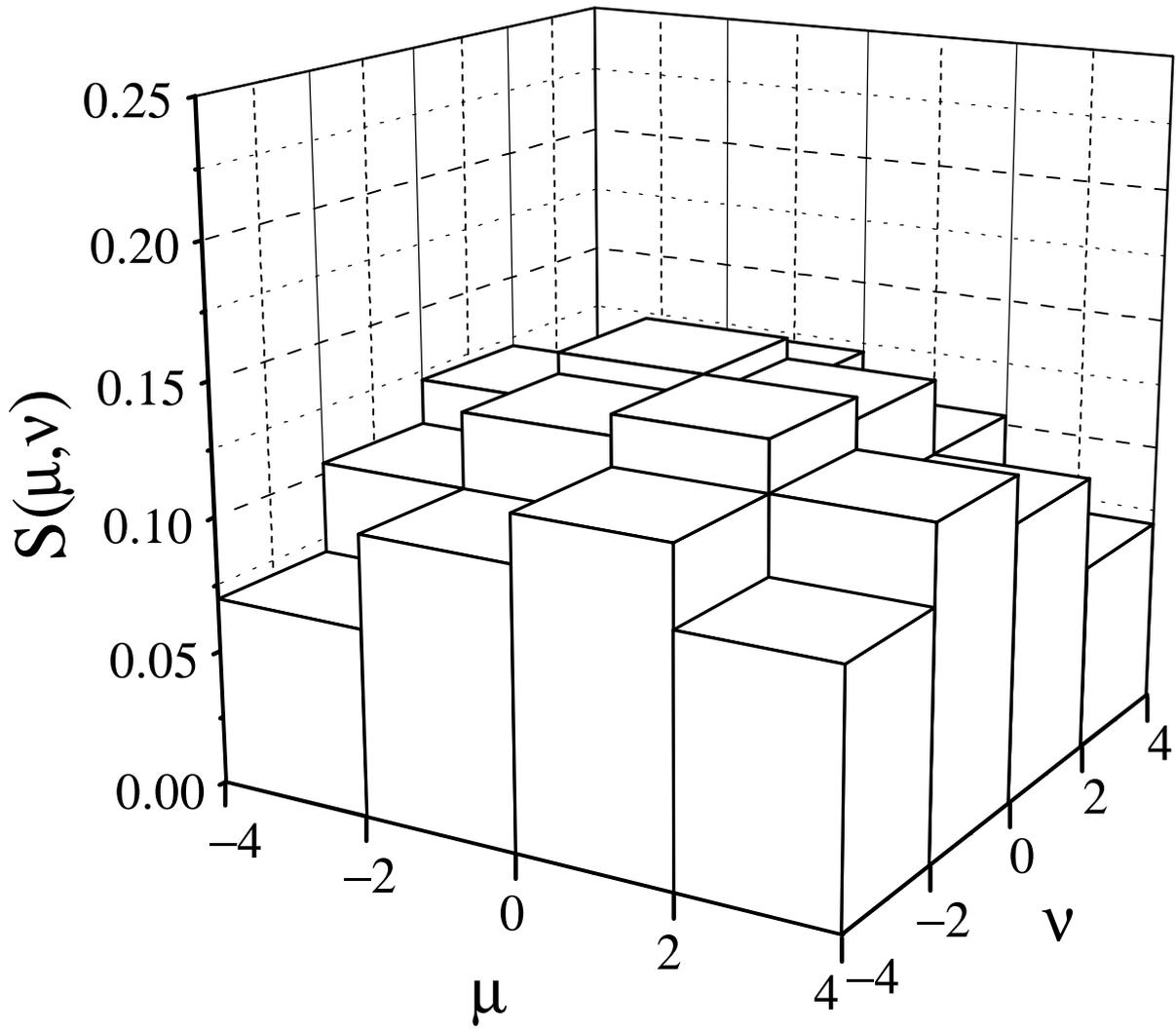
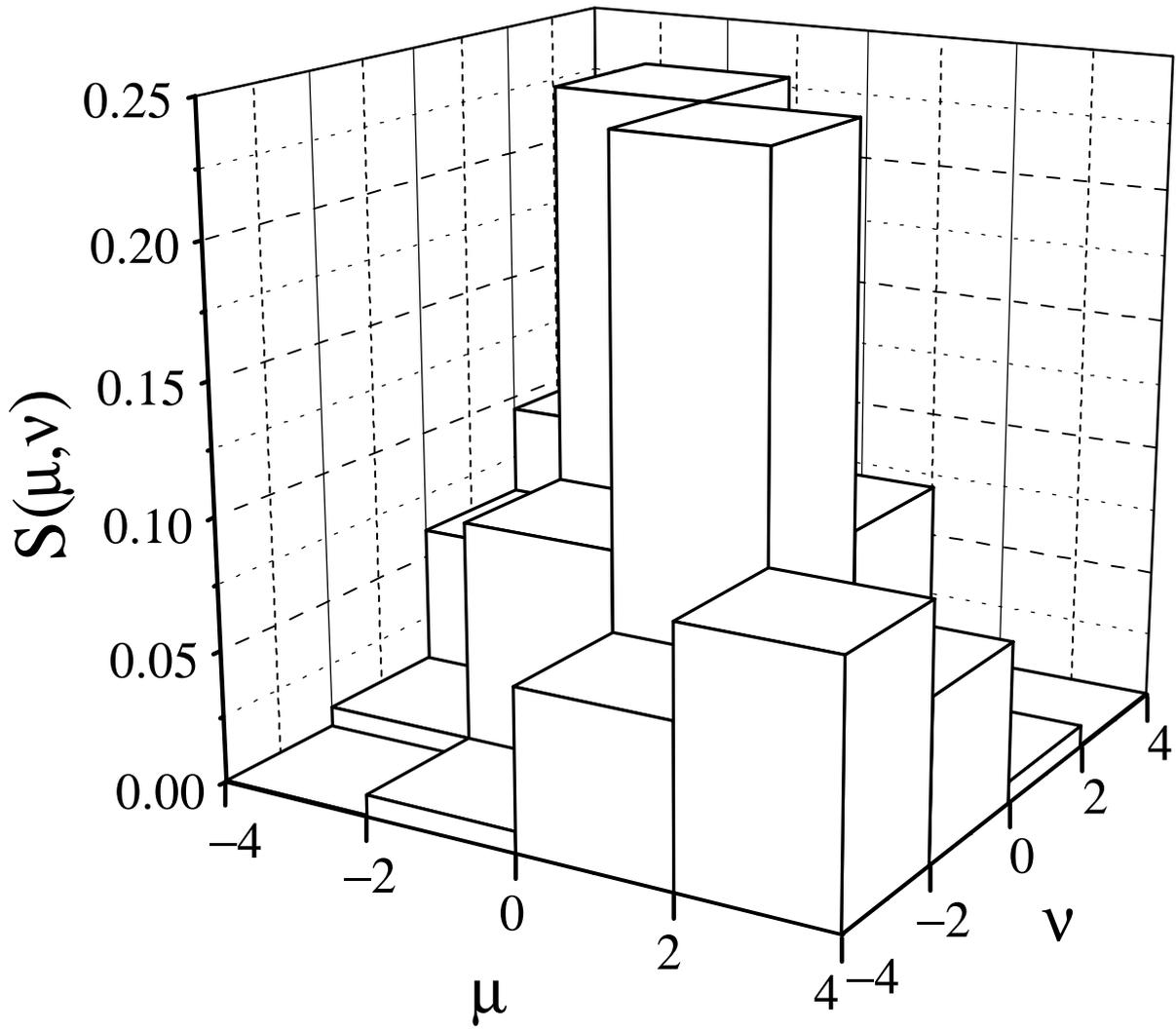
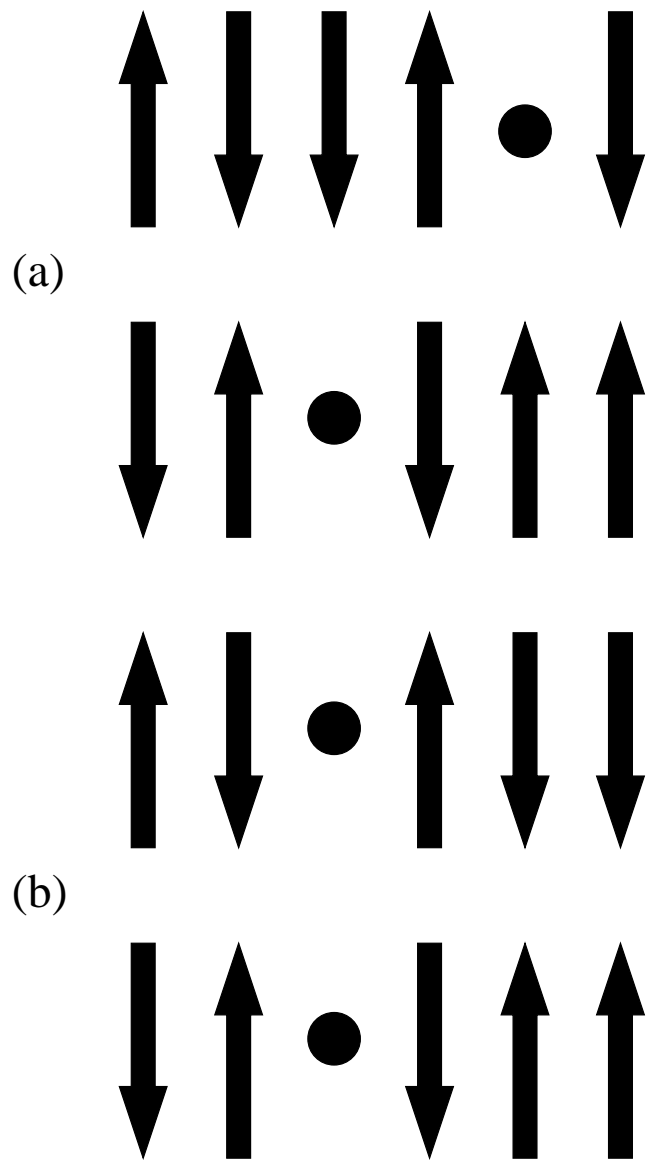
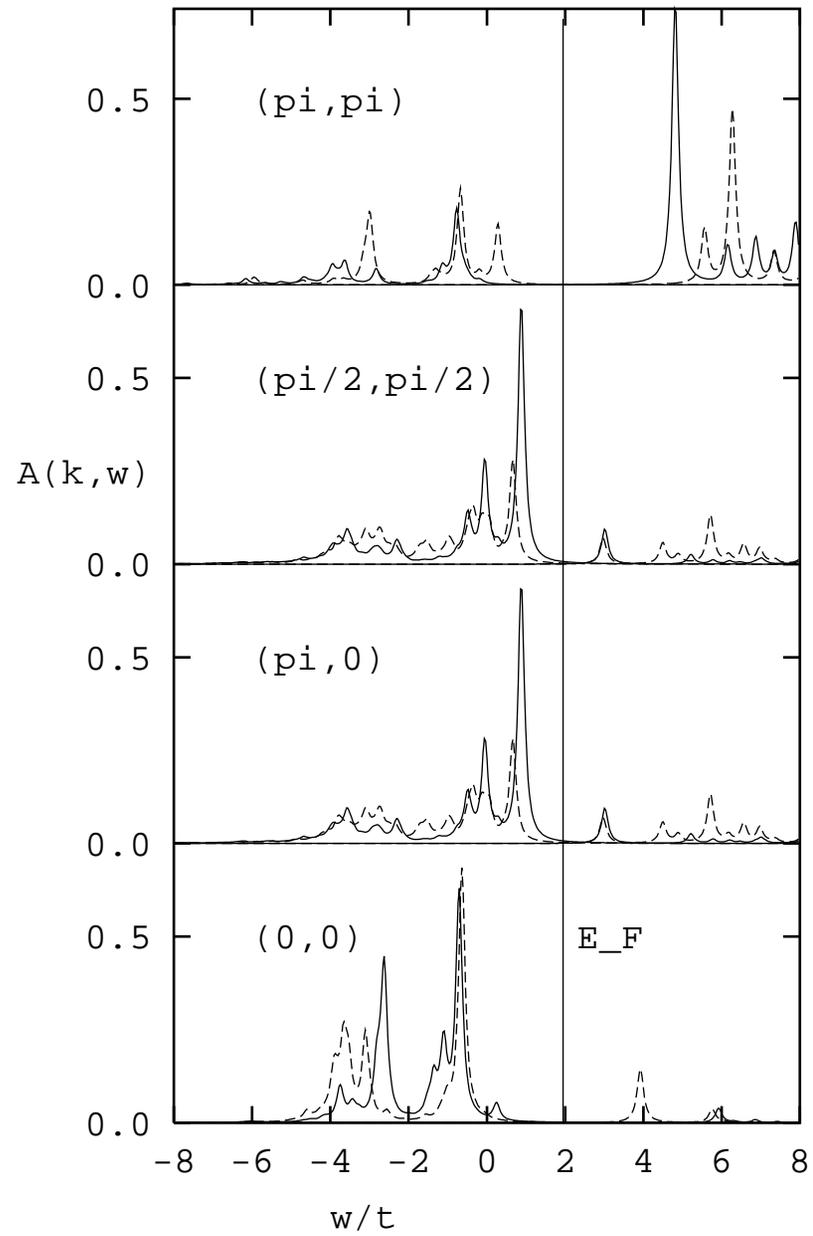
\end{document}